\documentclass[journal=jacsat,manuscript=article]{achemso}

\usepackage[version=3]{mhchem} 

\author{Paolo Biagioni}
\affiliation[CNISM - Dipartimento di Fisica, Politecnico di Milano]
{CNISM - Dipartimento di Fisica, Politecnico di Milano, Piazza Leonardo da Vinci 32, 20133 Milano, Italy}
\email{paolo.biagioni@polimi.it}
\phone{+39 02 2399 6599}
\fax{+39 02 2399 6126}

\author{Daniele Brida}
\affiliation[IFN - CNR - Dipartimento di Fisica, Politecnico di Milano]
{IFN - CNR - Dipartimento di Fisica, Politecnico di Milano,\\Piazza Leonardo da Vinci 32, 20133 Milano, Italy}

\author{Jer-Shing Huang}
\affiliation[Department of Chemistry, National Tsing Hua University]
{Department of Chemistry, National Tsing Hua University, Hsinchu 30013, Taiwan}

\author{Johannes Kern}
\affiliation[Experimentelle Physik 5, Universit\"{a}t W\"{u}rzburg]
{Nano-Optics and Bio-Photonics Group, Experimentelle Physik 5, Physikalisches Institut, Wilhelm-Conrad-R\"{o}ngten-Center for Complex Material Systems, Universit\"{a}t W\"{u}rzburg, Am Hubland, 97074 W\"{u}rzburg, Germany}

\author{Lamberto Du\`{o}}
\affiliation[CNISM - Dipartimento di Fisica, Politecnico di Milano]
{CNISM - Dipartimento di Fisica, Politecnico di Milano, Piazza Leonardo da Vinci 32, 20133 Milano, Italy}

\author{Bert Hecht}
\affiliation[Experimentelle Physik 5, Universit\"{a}t W\"{u}rzburg]
{Nano-Optics and Bio-Photonics Group, Experimentelle Physik 5, Physikalisches Institut, Wilhelm-Conrad-R\"{o}ngten-Center for Complex Material Systems, Universit\"{a}t W\"{u}rzburg, Am Hubland, 97074 W\"{u}rzburg, Germany}

\author{Marco Finazzi}
\affiliation[CNISM - Dipartimento di Fisica, Politecnico di Milano]
{CNISM - Dipartimento di Fisica, Politecnico di Milano, Piazza Leonardo da Vinci 32, 20133 Milano, Italy}

\author{Giulio Cerullo}
\affiliation[IFN - CNR - Dipartimento di Fisica, Politecnico di Milano]
{IFN - CNR - Dipartimento di Fisica, Politecnico di Milano,\\Piazza Leonardo da Vinci 32, 20133 Milano, Italy}

\title[An \textsf{achemso} demo]
  {Dynamics of four-photon photoluminescence in gold nanoantennas}


\begin{document}
\begin{abstract}
Two-pulse correlation is employed to investigate the temporal dynamics of both two-photon photoluminescence (2PPL) and four-photon photoluminescence (4PPL) in resonant and nonresonant nanoantennas excited at a wavelength of 800~nm. Our data are consistent with the same two-step model being the cause of both 4PPL and 2PPL, implying that the first excitation step in 4PPL is a three-photon $sp\rightarrow sp$ direct interband transition. Considering energy and parity conservation, we also explain why 4PPL behavior is favored over three-and five-photon photoluminescence in the power range below the damage threshold of our antennas. Since sizeable 4PPL requires larger peak intensities of the local field, we are able to select either 2PPL or 4PPL in the same gold nanoantennas by choosing a suitable laser pulse duration. We thus provide a first consistent model for the understanding of multiphoton photoluminescence generation in gold nanoantennas, opening new perspectives for applications ranging from the characterization of plasmonic resonances to biomedical imaging.

\end{abstract}

Noble-metal nanoparticles and nanoantennas are attracting an ever increasing attention since the resonant behavior
of the electron plasma oscillations in response to the external optical field enables numerous spectroscopic and sensing applications \cite{novotny}. The large near-field peak intensity enhancement that accompanies such resonances can give rise to a whole range of nonlinear optical effects.
In this context, two-photon photoluminescence (2PPL) imaging has become one of the preferred tools to study modal patterns of plasmonic resonances, especially  in gold \cite{Beversluis03,Schuck05,Muhlschlegel05,Ghenuche08,Huang10}. Two-photon absorption has also established gold nanoparticles among the most promising theragnostic biomarkers, since they provide at the same time the possibility for deep tissue imaging as well as for local heating in photothermal therapies \cite{Nagesha07,Durr07,Gobin07}.

At variance with coherent 2PPL in molecules, which usually proceeds through a virtual intermediate state, it has been proposed \cite{Imura} and experimentally demonstrated \cite{Biagioni} that 2PPL in gold is the result of two sequential single-photon absorption steps mediated by a real state. In detail, the first photon
excites an electron via an intraband transition within the $sp$ conduction band, while the second photon excites an electron from the $d$ band to recombine with an $sp$ hole in the conduction band. The dynamics of the TPPL signal, originated by the radiative recombination of $d$ holes, is ruled by the relaxation time of the transient distribution excited in the $sp$ conduction band after the first absorption event. Due to the increased density of states, interband radiative recombination in 2PPL occurs close to the L and X points in the reciprocal space, leading to two emission bands located in the green and red spectral regions, respectively \cite{Imura}. Higher-order absorption processes with characteristic power dependencies have also been reported in the literature. Three-photon absorption has been observed from single gold nanoparticles \cite{Farrer05}, while four-photon photoluminescence (4PPL) has been reported for resonant gold dipole antennas \cite{Muhlschlegel05}. Interestingly, in the case of 4PPL, an increased spectral weight of the green (L point) emission band was observed, compared to standard 2PPL \cite{Muhlschlegel05}. While the mechanism behind 2PPL is well understood, there is no consistent model for the observed higher-order nonlinearities and the appearance of either order in experiments seems to be elusively random.

Here we study the photoluminescence (PL) dynamics of both a rough gold film as well as high-quality resonant single-crystalline nanoantennas \cite{Huang10b}. While the rough gold film only shows 2PPL, for the nanoantennas we observe both 2PPL and 4PPL depending on the pulse length. In order to characterize the non-equilibrium dynamics of both 2PPL and 4PPL, we perform two-pulse correlation measurements using 100 fs pulses centered at 800~nm, in which PL is recorded as a function of the time delay between two pulse replicas \cite{Wang94,Lui10}. For both 2PPL and 4PPL, two-pulse correlation shows a single-exponential incoherent tail that directly probes the 1~ps relaxation time of the nonequilibrium electron distribution, which therefore represents the intermediate excited configuration for both cases. Based on this observation, we suggest that in the case of 4PPL the intermediate nonequilibrium electron distribution is created by a coherent three-photon transition mediated by virtual states, as implied by energy and parity conservation and supported by the experimental features of the 4PPL spectra. The observation of orders other than 2 and 4 can be excluded in the framework of this model for the 800~nm excitation wavelength. We provide further evidence for the validity of this picture by demonstrating control over the occurrence of 4PPL vs 2PPL from the same resonant nanostructure by changing the pulse duration for a given average power. In addition, we also find that the relaxation time of the hot electron distribution is shortened in the particular case of resonant optical antennas as compared to off-resonance antennas. Possible explanations for this intriguing behavior are discussed.

\begin{figure}
\includegraphics[width=0.78\textwidth]{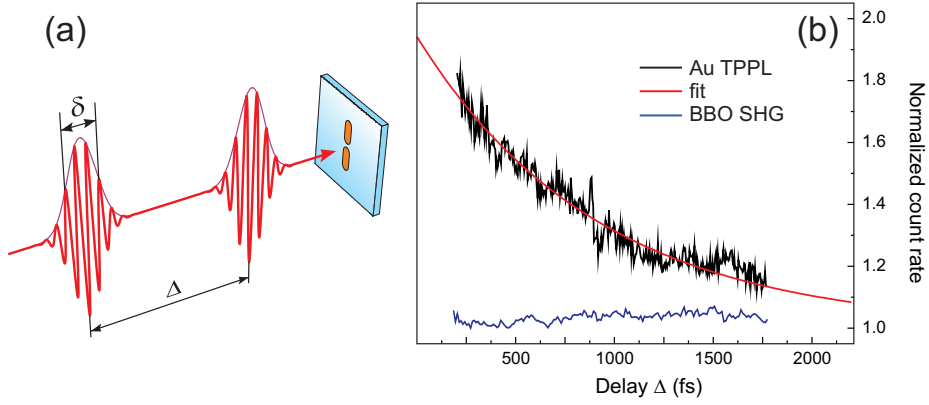}
\caption{\label{fig1} Panel (a): sketch of the two-pulse correlation measurements on  gold nanoantennas; panel (b): tail of the correlation curve (black line) for the 2PPL signal from a rough gold film. The red line is a best fit according to Eq.~2 (see text) with $\tau_{sp}$ as the only free parameter. The blue trace corresponds to the reference SHG correlation signal obtained from a BBO crystal placed at the sample position. The count rates are normalized to those obtained with two laser pulses separated by a time delay $\Delta\sim4$~ps.}
\end{figure}

The experimental setup is based on a commercial confocal microscope (Alpha-SNOM, WITec GmbH, Germany) coupled to a mode-locked Ti:sapphire oscillator producing 40~fs pulses at 100~MHz repetition rate and 800~nm wavelength. Collinear pulse pairs with computer-controlled delay $\Delta$ [see sketch in \ref{fig1}(a)] are generated by a balanced Michelson interferometer.
The output of the interferometer is coupled into a short piece of single-mode fiber, whose far end serves as a point-like source for the confocal microscope. Light is focused onto the sample by an oil-immersion objective with 1.4 numerical aperture. Precompensation of dispersion introduced by the optics in the beam path is achieved by a double pass in a pair of SF10 Brewster-cut prisms. After optimization of the precompensation stage, a pulse duration $\delta <100$ fs is obtained after the objective, as measured with second-harmonic generation (SHG) correlation from a thin $\beta$-barium borate (BBO) crystal.
Gold PL is collected using the same objective and sent to a photon counter through a multimode fiber whose 75~$\mu$m core also acts as a collection pinhole for background rejection. Suitable spectral filters (transmission window 450-750~nm) are inserted along the collection path in order to selectively detect the gold PL while completely rejecting both the fundamental excitation wavelength and possible second-harmonic signals.

In order to record a two-pulse PL correlation curve, the area of interest is first imaged by means of scanning confocal optical microscopy, then a selected structure is positioned in the focus and the PL signal is recorded as a function of the pulse delay $\Delta$. The typical power needed to obtain a sizable signal without damaging the sample was of the order of 100 $\mu$W per arm in the interferometer. Count rates are normalized to those obtained with two well-separated laser pulses (time delay $\Delta\sim4$~ps).
We generally observe that the 2PPL or 4PPL signals of continuously illuminated single- or multi-crystalline gold structures show intensity fluctuations, which are the main source of noise in the experiments.

We first investigate a reference sample for 2PPL, i.e., a rough polycrystalline gold film ($\sim 50$~nm thickness) fabricated by thermal evaporation \cite{Beversluis03}.
The corresponding 2PPL correlation trace is shown in \ref{fig1}(b) for delay times $\Delta>200$ fs, revealing a $\sim 1$~ps exponential tail. Comparison with the SHG correlation curve acquired replacing the gold film by a BBO crystal [blue line in \ref{fig1}(b)] confirms that this tail is indeed due to gold 2PPL dynamics and not to a residual chirp of the light pulses.

The observed $\sim 1$~ps dynamics is in full agreement with previous experimental results where the 2PPL yield was measured as a function of the pulse duration and the same limiting dynamics was observed \cite{Biagioni}. In order to understand this result, let us recall that gold 2PPL in the recorded spectral range is mostly generated by the recombination of holes in the 5$d$ band created via a two-step process involving two sequential one-photon absorption transitions \cite{Imura,Biagioni}. This excited hole distribution can relax radiatively and directly contribute to 2PPL, albeit with a small quantum yield. It can also recombine nonradiatively within the $sp$ band by generating localized plasmon oscillations that subsequently radiate, increasing the luminescence quantum yield \cite{Dulkeith,Wissert}.

The two sequential one-photon absorption events are governed by the following rate equations \cite{Biagioni}:
\begin{subequations}\label{eq:zero}
\begin{equation}\label{subeq:zeroA}
  \frac{dN_{sp}}{dt} = \sigma_{sp\rightarrow sp}NF(t) - \frac{N_{sp}}{\tau_{sp}} - \sigma_{d\rightarrow sp} N_{sp}F(t),
\end{equation}
\begin{equation}\label{subeq:zeroB}
  \frac{dN_{d}}{dt} = \sigma_{d\rightarrow sp} N_{sp}F(t) - \frac{N_{d}}{\tau_{d}},
\end{equation}
\end{subequations}
where $N$ is the sample density expressed as atoms per unit volume, $N_{sp}$ and $N_{d}$ are the densities of holes created below the Fermi level in the $sp$ and $d$ bands, respectively, while $\tau_{sp}$ and $\tau_{d}$ represent the
relaxation times of the $sp$ and $d$ holes. $F(t)$ describes the pulse intensity profile used for the excitation and $\sigma_{sp\rightarrow sp}$ ($\sigma_{d\rightarrow sp}$) is the cross section of the $sp\rightarrow sp$ ($d\rightarrow sp$) transition excited by the first (second) absorbed photon. The third term of the right-hand side of Eq.~1a is negligible in the small perturbation regime characterized by $N_{sp}\ll N$.

The 2PPL intensity $I_{\mathrm{2PPL}}(\Delta)$ can be obtained from Eq.~1 by calculating the temporal average value $\left\langle N_d \right\rangle$, to which the time-integrated photoluminescence signal is proportional \cite{Biagioni}. To do this, we consider a function $F(t)$ consisting of two identical light pulses of duration $\delta$ separated by a time delay $\Delta$. As already proposed in Ref.~\citenum{Biagioni}, the dynamics of the process is governed by $\tau_{sp}$, which is typically on the order of 1~ps in gold \cite{Biagioni,Sun,Groeneveld}. Since the  duration $\delta$ of the pulse impinging on the sample is below 100~fs, i.e. much shorter than $\tau_{sp}$, we obtain for the 2PPL intensity the following simple expression for delays $\Delta \gg \delta$:
\begin{equation}\label{eq:one}
   I_{\mathrm{2PPL}}(\Delta)\propto \left\langle N_{d}\right\rangle \propto I^{2}_0(1+e^{-{\frac{|\Delta|}{\tau_{sp}}}}),
\end{equation}
where $I_0$ is the time-averaged excitation intensity at the sample. Eq.~2 correctly reproduces the experimentally found exponential dependence of the 2PPL count rate on the time delay $\Delta$. The validity of Eq.~2 is independent of details of the laser pulses such as bandwidth and chirp, as long as $\delta \ll \tau_{sp}$ and $\delta \ll \Delta$. By fitting the experimental traces using Eq.~2 [red line in \ref{fig1}(b)] we find $\tau_{sp} \sim 910$~fs, in good agreement with the typical relaxation times observed for electrons and holes close to the Fermi level in gold films and nanostructures due to electron-phonon interactions \cite{Biagioni,Sun,Groeneveld}.

\begin{figure}
\includegraphics[width=0.68\textwidth]{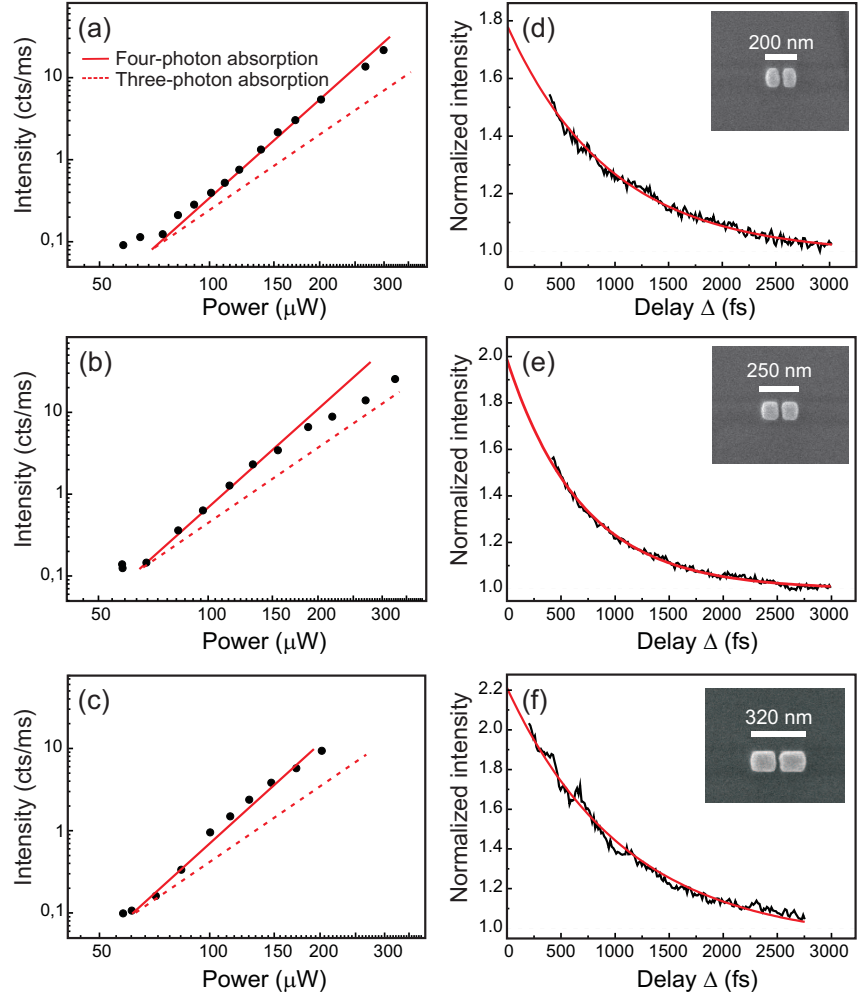}
\caption{\label{fig3} 4PPL power curves [panels (a), (b), and (c)] and normalized correlation curves [panels (d), (e), and (f)] for three different nanoantennas characterized by different arm lengths but an otherwise similar geometry. Red solid (dashed) lines in the power curves represent a guide for four-photon (three-photon) behavior. The correlation data are obtained with 100 $\mu$W average power per arm in the interferometer and then normalized to the count rate obtained with the two laser pulses separated by a time delay $\Delta\sim4$~ps. Insets are scanning electron microscopy images of the antennas. The red lines in panels (d), (e), and (f) are best fits obtained by using the function from Eq.~2 (see text).}
\end{figure}

We shall now discuss 4PPL correlation data obtained from gold nanoantennas. Antennas are prepared by focused ion-beam milling starting from single-crystalline gold flakes on an ITO-coated glass substrate \cite{Huang10b}. \ref{fig3}(a)-(c) reports the PL intensity as a function of the average excitation power for three gold dipole antennas with the same 10~nm gap and total lengths of 200, 250, and 320~nm, respectively [see SEM images in the insets of \ref{fig3}(d)-(f)]. The slope of the power curves in a log-log plot [\ref{fig3}(a)-(c)] clearly reveals a four-photon process \cite{Muhlschlegel05}. The 250~nm-long antenna shows a small deviation from slope of four for average powers higher than about 200~$\mu$W [\ref{fig3}(b)], which can be attributed to thermal effects, as we discuss below. Confocal 4PPL maps and finite-difference time-domain (FDTD) simulations show that the latter antenna is on resonance with the excitation wavelength. Indeed, in the experiment it provides the highest 4PPL count rate for a given excitation power in the confocal maps. Moreover, simulations show that, upon illumination at a wavelength of 800~nm with a diffraction-limited spot, it exhibits a local field intensity enhancement which is about a factor of 2 to 3 larger than those of the other two antennas. It is important to note that all antennas display single-exponential tails in the correlation curves [as shown in \ref{fig3}(d)-(f)] that are remarkably similar to those exhibited by the multi-crystalline gold film solely showing 2PPL [\ref{fig1}(b)].

This observation unravels a universal behavior of the dynamics of the 4PPL processes. In particular, it suggests that, irrespectively of the process order (two- or four-photon), the limiting intermediate step in the overall multi-photon cascade is always constituted by the same non-equilibrium Fermi distribution of $sp$ holes, which relaxes with a $\sim 1$~ps time constant, as already found for standard 2PPL in gold \cite{Biagioni}. The 4PPL correlation and power curves can be reproduced by the following model. We assume that the $d$ holes, whose radiative recombination produces the PL in a $n$-order process, are generated by a cascade of $n$ sequential absorption processes, each described by the following rate equation:
\begin{equation}\label{eq:two}
   \frac{dN_{i}}{dt} = \sigma_{i}N_{i-1}(t)F(t) - \frac{N_{i}(t)}{\tau_{i}},
\end{equation}
where $\tau_i$ and $N_i$ are the relaxation time and the density of holes in the $i^{th}$ intermediate configuration, respectively, and $\sigma_i$ is the cross section of the transition from the $(i-1)^{\mathrm{th}}$ to the $i^{\mathrm{th}}$ configuration. The initial conditions are $N_0(0) = N$ and $N_i(0)=0$ for $i=1...n$. By solving this system of coupled equations in the $\delta\ll\Delta$ limit, one obtains
\begin{equation}\label{eq:three}
   I_{\mathrm{MPPL}}(\Delta)\propto \left\langle N_{d}\right\rangle=\left\langle N_{n}\right\rangle \propto I_0^n (1+\overset{n-1}{\underset{i=1}{\sum }} e^{-{\frac{|\Delta|}{\tau_{i}}}}).
\end{equation}
Here, each of the $n$ steps can in principle be related to a number of events, including interband transitions \cite{Mooradian}, intraband transitions \cite{Beversluis03}, or generation/annihiliation of plasmons \cite{Dulkeith}. In particular, nonlinear excitation of a $sp$ hole can account for the additional steps leading from 2PPL to 4PPL. Such processes are well-known and currently employed in multi-photon photoemission experiments \cite{Montaut95} and as a probe in photoemission electron microscopy to investigate local field enhancements in noble-metal nanostructures \cite{aeschlimann,Douillard08}. Remarkably, this model holds even when some of the transitions involve a ``virtual'' intermediate state, i.e. when transitions in the cascade are excited by a coherent multi-photon absorption. In this case the corresponding relaxation time $\tau_{i}$ in Eq.~3 is equal to zero.

For $n=4$, Eq.~4 correctly reproduces the observed 4PPL power dependence, provided $\tau_i \sim 0$ is assumed for three of the four steps. The single-exponential correlation tail observed also in the case of 4PPL, when compared to Eq.~4, thus suggests that the dynamics is still dominated by an intermediate distribution having a relaxation time of the order of $\sim1$~ps and that all other contributions leading to $n=4$ are characterized by a relaxation time $\tau_i$ much shorter than the investigated time window (i.e. shorter than about 100~fs).

\begin{figure}
\includegraphics[width=0.9\textwidth]{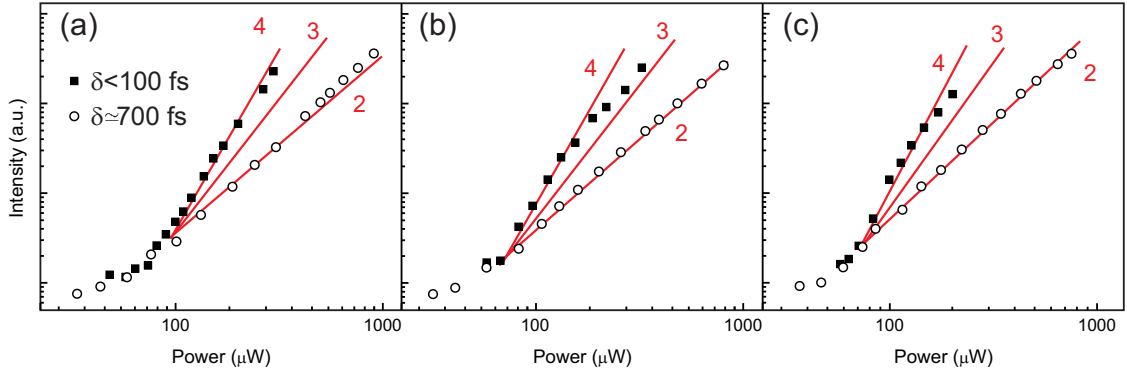}
\caption{\label{fig4} Comparison between the multi-photon emission curves for the three investigated antennas acquired with pulse durations $\delta <100$~fs and $\delta > 700$~fs. The antenna length is a) 200~nm, b) 250~nm, c) 320~nm. Lines with slopes 2, 3, and 4 in the log-log plot are also shown as a guide for the eye.}
\end{figure}

Our model therefore confirms that the dynamics can be very similar for different multi-photon absorption orders since the same longer-lived intermediate state dominates the absorption cascade, whose order $n$ is in any case revealed by the dependence on the average excitation intensity [\ref{fig3}(a)-(c)]. The dynamics of the other absorption steps leading to $n>2$ in 4PPL is confined within the first 100~fs that are not accessed in our investigation. Based on this interpretation, it is therefore expected that the yield of 4PPL strongly depends on the excitation pulse duration, as opposed to standard 2PPL \cite{Biagioni}. In order to verify this prediction, we repeat the measurement of the absorption order in the investigated antennas by sending fs pulses through a longer piece of optical fiber, leading to a pulse duration $\delta > 700$~fs. Remarkably, we observe a transition from 4PPL to 2PPL for the same antennas (see \ref{fig4}).

\begin{figure}
\includegraphics[width=0.55\textwidth]{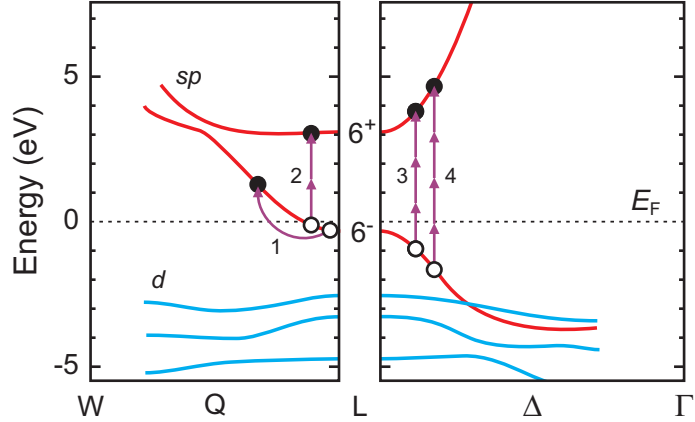}
\caption{\label{fig5} Gold relativistic band structure near the L point of the reciprocal space according to Ref.~\citenum{Eckardt}. Multi-photon direct $sp\rightarrow sp$ transitions are indicated by the vertical arrows. A bent arrow shows one of the possible indirect single-photon transitions. At the L point, the $sp$-derived states above and below the Fermi level respectively belong to the $6^+$ (even) and $6^-$ (odd) irreducible representations of the D$_\textit{6d}$ double group describing the symmetry properties of the reciprocal space at L. }
\end{figure}

In the following we propose a consistent picture to describe the absorption steps leading to 4PPL, which also explains why 4PPL excited with 800 nm wavelength dominates over alternative multi-photon processes. As a starting point, let us consider the first absorption step, which in the case of 2PPL is an intraband indirect transition within the $sp$ band. We further assume that in the case of 4PPL this $sp\rightarrow sp$ transition involves absorption of more than one photon. In this case, it is reasonable to assume that this multi-photon transition will be vertical in the reciprocal space, since a cascade of indirect absorption events would be exceedingly unlikely. By close inspection of the gold band structure \cite{Eckardt} in the proximity of the L and X points, it can be seen that the direct gap between adjacent $sp$ bands is about 3~eV around the L point (see \ref{fig5}) and  about 6.5~eV around the X point. The onset of the joint density of states (JDOS) in gold is therefore near the L point of the reciprocal space. Since the $sp$ bands above and below the Fermi level at L disperse very rapidly, direct transitions can occur only within a small volume of the reciprocal space close to L. Such energy-allowed transitions, involving 2, 3, or 4 photons, are sketched in \ref{fig5} for the photon energy employed in our experiments (1.55 eV), together with a possible indirect one-photon transition typical for 2PPL. It should however be noted that the two $sp$ bands drawn in the figure have different parity character at the L point, and that multi-photon transitions couple states with the same (opposite) parity when the number of photons involved in the process is even (odd). Therefore, parity conservation suggests that three-photon $sp\rightarrow sp$ transitions are likely to be privileged with respect to two- or four-photon transitions \cite{L}. Moreover, two-photon transitions occur at the very onset of the JDOS (along the L-W direction) and are therefore further hindered. Also, by inspection of the Au bands at L, one can immediately see that the energy differences corresponding to three- or four-photon $sp\rightarrow sp$ transitions are higher than the energy of the Van Hove singularity at L, which corresponds to a local maximum in the JDOS. Therefore, the JDOS has to be lower for four-photon compared to three-photon transitions, adding further consistency to the hypothesis that four-photon transitions should be much less effective than three-photon transitions. This leads to the conclusion that a three-photon$sp\rightarrow sp$ transition followed by a single-photon $d\rightarrow sp$ absorption resulting in 4PPL should be much more efficient than other multi-photon processes leading e.g. to three- or five-photon photoluminescence. Incidentally, we note here that even if a two-photon $sp\rightarrow sp$ transition would occur, this would create an $sp$ hole close to the Fermi level (see \ref{fig5}). Since the $d$ band lies more than 2~eV below the Fermi level, this would require a two-photon $d\rightarrow sp$ transition to fill the energy distance between the $d$ and the $sp$ band, thus resulting again in 4PPL.

This proposed model is able to explain all experimental observations, including previous studies, concerning 4PPL from gold nanostructures: (i) the three-photon $sp\rightarrow sp$ absorption step occurs through virtual levels ($\tau_i=0$) and is thus an instantaneous process, so that the relaxation time of the $sp$ hole distribution solely determines the dynamics of the 4PPL process. The 2PPL and 4PPL two-pulse correlation experiments should thus display the same temporal behavior, as indeed observed in the experiments. (ii) The initial instantaneous three-photon absorption process in 4PPL should display a strong intensity dependence and its yield, for a constant average excitation power, should scale as $1/{\delta^2}$: therefore lengthening the excitation pulsewidth from 100 to 700~fs should reduce the 4PPL yield by a factor of $\sim50$. This is in agreement with the disappearance of 4PPL for the longer excitation pulses. (iii) At variance with 2PPL, 4PPL is the result of recombination processes occurring mainly at the L point of the reciprocal space. This is indeed fully consistent with the already mentioned increased spectral weight of the green emission (a signature of radiative $d\rightarrow sp$ recombination close to L) that was reported in Ref.~\citenum{Muhlschlegel05} for antennas beyond the threshold for 4PPL.

These arguments do not completely rule out the possibility of observing PL power curves with intermediate slopes between those characteristic for 2PPL and 4PPL, as reported in the literature \cite{Farrer05,Wissert}. Such an intermediate behavior could be interpreted either as due to illumination conditions (power, pulse duration) exciting both 2PPL and 4PPL with similar efficiencies, or as a consequence of thermal effects \cite{Rotenberg,Perner,Pelton}, which can cause saturation of the empty states above the Fermi level and increase electron-electron scattering, leading in both cases to a reduction of the PL yield.

Closer inspection of the fitting results for the correlation traces in \ref{fig3} reveals that the 250~nm resonant antenna actually displays a shorter relaxation time compared to the non-resonant ones. Indeed, the 200~nm and 320~nm antennas provide $\tau_{sp}\sim 1000\pm100$~fs, in very good agreement with the results obtained for 2PPL in gold and with typical electron-phonon relaxation times, while for the resonant antenna we find $\tau_{sp}\sim 650\pm50$~fs. Other measurements on non-resonant gold nanostructures all provide relaxation times similar to those found for 2PPL. Here we offer two possible explanations for this peculiar behavior:

(i) because of the more efficient excitation of the resonant antenna, $\tau_{sp}$ might be reduced by thermal effects, leading to smearing of the Fermi distribution function and thermally induced damping of the plasmon oscillation \cite{Rotenberg,Perner,Pelton}, in agreement with the observed slight deviation from a pure four-photon behavior in \ref{fig3}(b);

(ii) a second intriguing possibility is that the intermediate excited $sp$-band electron distribution, required for 4PPL, efficiently decays into the resonant antenna plasmon modes \cite{Dulkeith}. This would represent a further loss channel directly contributing to a decrease in $\tau_{sp}$. In order to gain a qualitative insight into this possibility, we have run FDTD simulations to calculate the enhancement in the total decay rate from an emitting dipole in the 800-1000 nm range (simulating intraband recombination)\cite{Beversluis03} coupled to the antenna feedgap. We find that for the resonant antenna the enhancement in the decay rate (i.e. the antenna Purcell factor) is a factor of $\sim 10$ larger than for the two non-resonant antennas therefore providing a possible explanation for the differences in $\tau_{sp}$ between resonant and off-resonance antennas.

In conclusion, we have analyzed the multi-photon photoluminescence dynamics in gap nanoantennas by performing two-pulse correlation experiments. By exploiting the large near-field intensity enhancement afforded by the antennas we have observed 4PPL, for which the temporal dynamics is dominated by the same $\sim 1$~ps relaxation time as in the case of 2PPL. The two processes hold therefore strong similarities, since their dynamics is always limited by carrier relaxation in the $sp$ band, and correspondingly they proceed via the same intermediate configuration. Further absorption steps that increase the process order from $n=2$ to $n=4$ all happen within the first 100~fs, i.e. at time scales shorter than the investigated time window, and as such they are more sensitive to the actual pulse duration in the sub ps regime. Indeed, we also demonstrate that, for the same nanoantennas, 2PPL is obtained instead of 4PPL when the pulse duration is increased from $\delta<100$~fs to $\delta\sim700$~fs for the same average power. Therefore the 4PPL yield in gold, as opposed to the yield of 2PPL \cite{Biagioni}, is strongly dependent on the pulse duration used for excitation, providing a handle for switching between 2PPL and 4PPL. High-quality single-crystalline nanoantennas were needed for the investigation since they reliably provide the local field enhancement in order to boost such nonlinear processes \cite{Huang10b}. Based on on our observations, we provide a first consistent framework for the explanation of 4PPL, where a three-photon direct transition between different $sp$ bands is followed by a one-photon transition between the $d$ and $sp$ bands. Interestingly, a shortened relaxation time has been found for the resonant nanoantenna, which is attributed either to thermal effects or to the resonant decay of the intermediate excited distribution into the plasmon antenna mode.

\begin{acknowledgement}

 The authors thank M. Savoini for insightful discussions and for his help during the experiments. Financial support from DFG (SPP 1391 ``Ultrafast Nano-optics''), from National Science Council of Taiwan (NSC 99-2113-M-007-020-MY2), from Fondazione Cariplo (``Engineering of optical nonlinearities in plasmonic metamaterials''), from MIUR (PRIN project No. 2008J858Y7), and from the European Union Nano Sci-European Research Associates (``FENOMENA'') is gratefully acknowledged.

\end{acknowledgement}





\begin{thebibliography}{99}
\bibitem{L}
Parity is a good quantum number only for the electron states at L, nevertheless two-, three-, and four photon transitions can occur only very close to this point, where states are likely to maintain a well-defined parity.
\end{thebibliography}
\end{document}